\documentclass[letter,twocolumn]{jpsj3}
\usepackage{txfonts}

\usepackage{graphicx}
\usepackage{dcolumn}
\usepackage{bm}
\usepackage{color}
\usepackage{ulem}
\usepackage{comment}

\newcommand{\dg}{\dagger}

\title{
Phase Competition and Superconductivity in $\kappa$-(BEDT-TTF)$_2$X:\\
Importance of Intermolecular Coulomb Interactions
}

\author{Hiroshi Watanabe$^{1}$\thanks{h.watanabe@aoni.waseda.jp}, Hitoshi Seo$^{2,3}$, and Seiji Yunoki$^{2,3,4}$}
\inst{
$^1$Waseda Institute for Advanced Study, Waseda University, Shinjuku, Tokyo 169-8050, Japan\\
$^2$RIKEN, Wako, Saitama 351-0198, Japan\\
$^3$RIKEN Center for Emergent Matter Science (CEMS), Wako, Saitama 351-0198, Japan\\
$^4$RIKEN Advanced Institute for Computational Science (AICS), Kobe, Hyogo 650-0047, Japan
}

\abst{
We theoretically study the competition among different electronic phases in molecular conductors $\kappa$-(BEDT-TTF)$_2$X.
The ground-state properties of a 3/4-filled extended Hubbard model with the $\kappa$-type geometry are investigated by a variational Monte Carlo method.  
We find various competing phases: dimer-Mott insulator, polar charge-ordered insulator, 3-fold charge-ordered metal, and superconductivity,
whose pairing symmetry is an ``extended-$s$+$d_{x^2-y^2}$''-wave type.
Our results show that the superconducting phase is stabilized not on the verge of the Mott metal-insulator transition but near charge order instabilities,
clearly indicating the importance of the intradimer charge degree of freedom and the intermolecular
Coulomb interactions, beyond the simple description of the half-filled Hubbard model.
}

\begin{document}
\maketitle

Metal-insulator (MI) transition and related emergent phenomena in strongly correlated electron systems
such as transition metal oxides~\cite{Imada} and molecular conductors,~\cite{Lebed} are under extensive study.
The most typical example is the Mott transition, around which various phases are induced by tuning controlling parameters,
most prominently, the high-$T_c$ superconductivity (SC) in copper oxides~\cite{Bednorz} and the colossal magnetoresistance in manganese oxides,~\cite{Chahara,Helmolt}
both achieved by changing the band filling from Mott insulators by chemical substitution.

The family of quasi two-dimensional molecular conductors $\kappa$-(BEDT-TTF)$_2$X, abbreviated as $\kappa$-(ET)$_2$X, has also been widely studied as a platform of Mott physics.~\cite{Kagawa,Kanoda}
In contrast with the cases above, the MI transition in $\kappa$-(ET)$_2$X is realized by bandwidth control, tuned by chemical (substitution of X) and/or physical pressure.
The insulating state is characterized by carrier holes localized on every dimer of ET molecules, then called the dimer-Mott insulator (DMI).
Depending on X, the degree of geometrical frustration on the localized spins is varied, and either antiferromagnetic (AF) order or spin-liquid behavior is observed at low temperatures.
On the other hand, SC appears almost ubiquitously in the metallic side, from the border of the MI transition.~\cite{Ardavan}
The $\kappa$-(ET)$_2$X system raises several important issues in condensed matter physics:
What are the effects of geometrical frustration on MI transition, magnetism, and SC? 
How far does the analogy between high-$T_c$ cuprates and molecular Mott systems, often discussed,~\cite{McKenzie} hold?

The theoretical model most intensively studied for $\kappa$-(ET)$_2$X is the dimer Hubbard model, i.e., the half-filled Hubbard model on the anisotropic triangular lattice.~\cite{Kino1}
There, the dimers of ET molecules facing each other are regarded as lattice sites.
It is a fundamental model to study the Mott transition, and successfully describes many aspects of the MI transition in $\kappa$-(ET)$_2$X.
However, despite extensive studies,~\cite{Kino2,Kondo,Schmalian,Kyung,Morita,Koretsune1,TWatanabe,Shinaoka,Dayal,Tocchio1,Laubach,Shirakawa}
it is controversial whether the model can properly describe the spin-liquid phase and SC in $\kappa$-(ET)$_2$X.
It is worth noting that several recent works have suggested that there is no region of SC in the entire parameter space.~\cite{TWatanabe,Dayal,Tocchio1}

Originally, the dimer Hubbard model is derived from the 3/4-filled Hubbard model where the basis functions are molecular orbitals of ET molecules,~\cite{Seo1,Seo2}
in the limit of strong dimerization and a large intramolecular (on-site) Coulomb interaction (the dimer approximation).~\cite{Kino1}
Some experimental studies~\cite{Manna,Abdel-Jawad,Guterding1} suggest that the intradimer charge degree of freedom is active, which is discarded in the dimer Hubbard model.
Indeed, theoretical studies~\cite{Kino1,Seo3,Kuroki1,Hotta,Naka1,Sekine,Guterding2} have been carried out based on 3/4-filled models for $\kappa$-(ET)$_2$X,
but they are limited to theoretical treatments where the interplay between the MI transition induced by strong correlation and SC cannot be properly treated.

In this paper, we study the 3/4-filled extended Hubbard model (EHM) for $\kappa$-(ET)$_2$X where the intradimer charge degree of freedom and intermolecular (intersite) Coulomb interactions are taken into account.
We investigate the phase competition and the mechanism of SC using a variational Monte Carlo (VMC) method,~\cite{McMillan,Ceperley,Yokoyama}
which enables us to study the MI transition and the associated emergent states with accuracy, including the quantum fluctuation.
We determine the ground-state phase diagram containing several spin- and charge-ordered phases and SC, which is characterized by an ``extended-$s$+$d_{x^2-y^2}$''-wave symmetry.
We show the importance of the intersite Coulomb interactions and discuss the mechanism of SC beyond the picture of the dimer Hubbard model. 

\begin{figure}[t]
\begin{center}
\includegraphics[width=7cm]{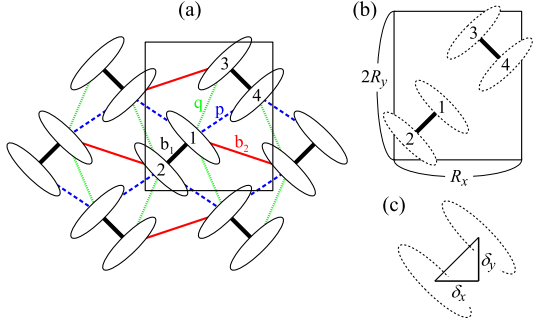}
\caption{\label{fig1} 
(Color online) Two-dimensional lattice structure for $\kappa$-(ET)$_2$X.
(a) Overview, (b) unit cell, and (c) configuration of a dimer pair.
The unit cell contains four molecules labeled as 1--4.
They are connected with bonds $b_1$, $b_2$, $p$, and $q$.
The dimers (1-2 and 3-4) form an anisotropic triangular lattice. 
}
\end{center}
\end{figure}

Following Refs. 10 and 28,
the model studied here is described by the Hamiltonian
\begin{equation}
H=-\sum_{\left<i,j\right>\sigma}t_{ij}(c^{\dagger}_{i\sigma}c_{j\sigma}+\text{H.c.})
+U\sum_{i}n_{i\uparrow}n_{i\downarrow}
+\sum_{\left<i,j\right>} V_{ij}n_in_j,
\end{equation}
where $c^{\dg}_{i\sigma}$ ($c_{i\sigma}$) is a creation (annihilation) operator of electron at the molecular site $i$ with spin $\sigma\,(=\uparrow,\downarrow)$,
$n_{i\sigma}=c^{\dagger}_{i\sigma}c_{i\sigma}$ and $n_i=n_{i\uparrow}+n_{i\downarrow}$.
$U$ and $V_{ij}$ are on-site and intersite Coulomb repulsions, respectively.
$\left<i,j\right>$ denotes a pair of neighboring molecules in the $\kappa$-type geometry,
labeled by $b_1$, $b_2$, $p$, and $q$, as shown in Fig.~\ref{fig1}(a).

We use tight-binding parameters $t_{ij}$ for deuterated $\kappa$-(ET)$_2$Cu[N(CN)$_2$]Br, which locates very close to the MI transition,~\cite{Miyagawa}
adopted from a first-principles band calculation as
$(t_{b_1},\,t_{b_2},\,t_p,\,t_q) = (196,\,65,\,105,\,-39)\,\text{meV} = (1.0,\,0.332,\,0.536,\,-0.199)\,t_{b_1}$.~\cite{Koretsune2}
We set the largest hopping integral $t_{b_1}$ as a unit of energy in the following.
When the dimer approximation is applied, the anisotropy in the hopping integrals is estimated as $\sim 0.5$.~\cite{Koretsune2}
Therefore, the degree of geometrical spin frustration is expected to be weak in this compound.

In the calculations below, we consider explicitly the intersite distances between molecular pairs.
In order to estimate them, we describe the lattice structure of $\kappa$-(ET)$_2$X by setting the parameters $(R_x,R_y,\delta_x,\delta_y)$, as shown in Figs.~\ref{fig1}(b) and~\ref{fig1}(c).
The unit cell is a rectangle with $R_x\times 2R_y$ and $\bm{\delta}=(\delta_x,\delta_y)$ is a vector connecting the centers of a dimer pair.
Here, we set $(R_x,R_y,\delta_x,\delta_y)=(1.0,0.7,0.3,0.3)\,R_x$ with $R_x$ as a unit of length.~\cite{CCDC}

The effect of Coulomb interactions is treated using a VMC method. 
We consider the trial wave function
$\left|\Psi\right>=P_{\text{J}_{\text{c}}}P_{\text{J}_{\text{s}}}
\left|\Phi\right>$,
where $\left|\Phi\right>$ is a Slater determinant constructed by diagonalizing the one-body Hamiltonian including the off-diagonal elements 
$\{D\}$, $\{M\}$, and $\{\Delta\}$, which induce long-range ordering of charge, spin, and SC, respectively (See Supplemental Material~\cite{sm} for the explicit form).
The renormalized hopping integrals are also included in $\left|\Phi\right>$ as variational parameters, ($\tilde{t}_{b_1}, \tilde{t}_{b_2}, \tilde{t}_p, \tilde{t}_q$),
where $\tilde t_{b_1}=t_{b_1}$ is fixed as a unit.
$P_{\text{J}_{\text{c}}}=\exp[-\sum_{i,j}v^{\text{c}}_{ij}n_in_j]$ and
$P_{\text{J}_{\text{s}}}=\exp[-\sum_{i,j}v^{\text{s}}_{ij}s^z_is^z_j]$ are
charge and spin Jastrow factors that control long-range charge and spin correlations, respectively.
Here, $v^{\text{c}}_{ij}=v^{\text{c}}(r_{ij})$ and $v^{\text{s}}_{ij}=v^{\text{s}}(r_{ij})$ are assumed, where $r_{ij}=|\bm{r}_i-\bm{r}_j|$ and $\bm{r}_i$ is the position of the molecular site $i$.
The variational parameters in $\left|\Psi\right>$ are therefore 
$\tilde{t}_{b_2}$, $\tilde{t}_p$, $\tilde{t}_q$, $\{D\}$, $\{M\}$, $\{\Delta\}$, $\{v^{\text{c}}_{ij}\}$, and $\{v^{\text{s}}_{ij}\}$,
and they are simultaneously optimized using the stochastic reconfiguration method.~\cite{Sorella}
The total number of molecular sites is $4\times L\times L/2=2L^2$ and varied from $L=12$ to $L=24$ with antiperiodic boundary conditions in both directions of the primitive lattice vectors.

\begin{figure}
\begin{center}
\includegraphics[width=7.5cm]{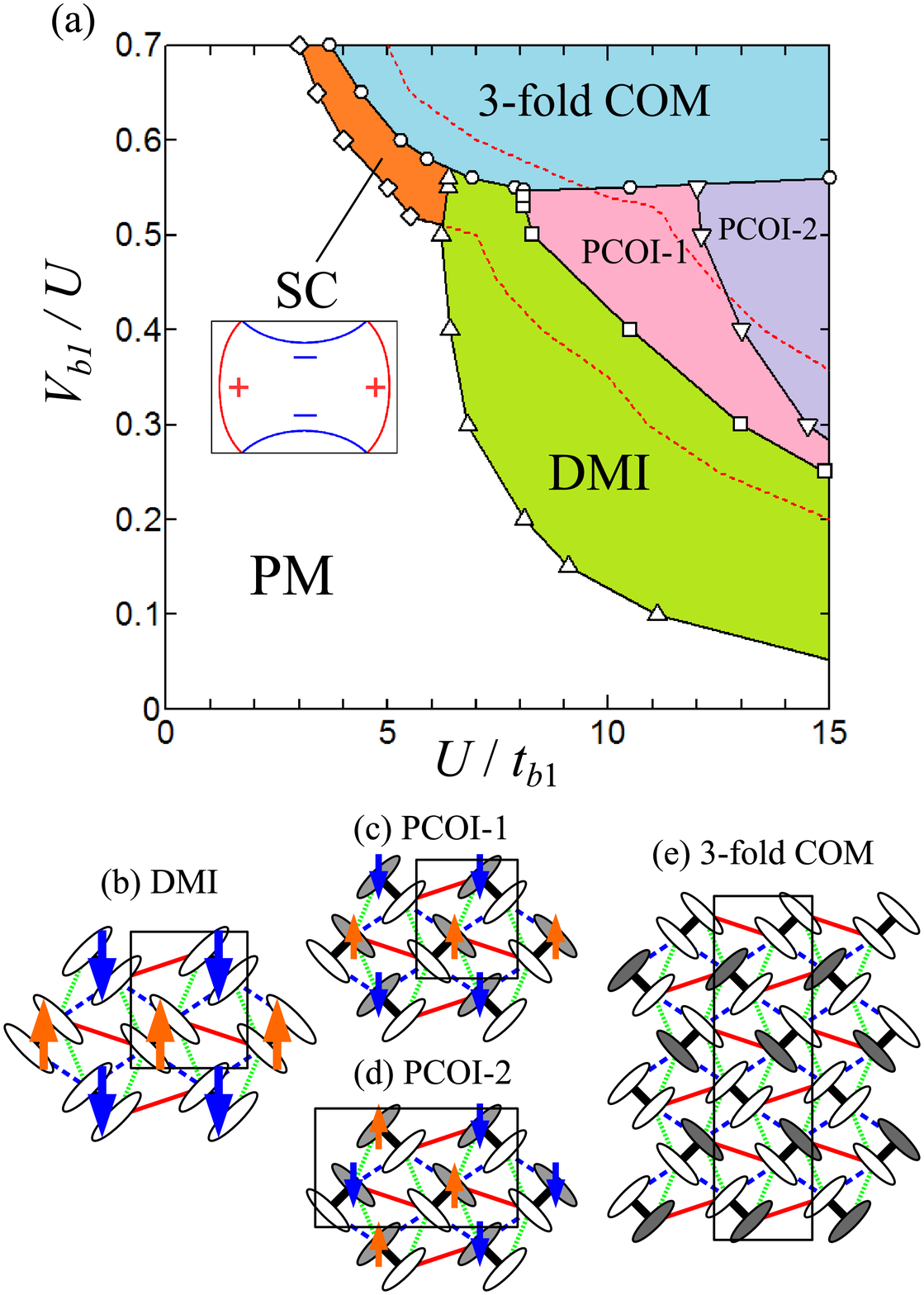}
\caption{\label{fig2}
(Color online) (a) Ground-state phase diagram of the 3/4-filled EHM for $\kappa$-(ET)$_2$X. 
The red dotted curves indicate the region where the SC condensation energy is finite. 
The SC gap structure on the Fermi surface is also indicated in the inset. 
The phase boundaries are determined by the calculations for $L=24$.
(b)-(e) Schematic view of each long-range ordered phase.
Up (down) arrows in (b)-(d) represent up (down) spin-rich sites.
In PCOI states (c) and (d), spin-rich sites are hole-rich sites (solid ovals) as well,
while in the DMI state (b), the hole density on each site is uniform. 
Solid (open) ovals represent hole-rich (hole-poor) sites in (e). 
The black rectangles in (b)-(e) are unit cells.
}
\end{center}
\end{figure}

Figure~\ref{fig2}(a) summarizes the ground-state phase diagram of the 3/4-filled EHM for $\kappa$-(ET)$_2$X with $L=24$ (1,152 molecular sites).
The on-site Coulomb interaction $U/t_{b_1}$ and the ratio between $U$ and the largest intersite Coulomb interaction $V_{b_1}/U$ are varied as parameters. 
The other intersite Coulomb interactions are set as $(V_{b_2},\,V_p,\,V_q)=(0.56,\,0.66,\,0.58)\,V_{b_1}$,~\cite{note1} assuming the $1/r$ dependence, whereas further long-range terms are neglected.
We find six distinct phases in the phase diagram: paramagnetic metal (PM), DMI with AF spin order, two charge-ordered insulating phases, 3-fold charge-ordered metal (3-fold COM), and SC.
The former charge ordering breaks the inversion symmetry, thus called polar charge-ordered insulator, with different AF spin orders (PCOI-1 and PCOI-2).~\cite{Naka1}
 
First, note that no long-range-ordered phases appear for $V_{b_1}/U\sim 0$ even when $U/t_{b_1}$ is large.~\cite{DeSilva}
This is different from the mean-field results, where $U$ stabilizes the AF DMI phase for $U/t_{b_1} \gtrsim 3$.~\cite{Kino1}
As $V_{b_1}/U$ increases, electrons begin to localize and DMI appears: A Mott MI transition occurs at finite $V_{b_1}/U$.
This result can be understood when we consider the effective ``on-dimer'' Coulomb repulsion $U_{\rm dim}$ discussed in the dimer approximation.
Without $V_{b_1}$, $U_\textrm{dim} = 2t_{b_1} - 4t_{b_1}^2/U + ... $,~\cite{Kino1} which cannot exceed the critical value for the Mott transition in our parameter set;
The dimerization is not sufficiently strong for realizing DMI. 
However, $U_\textrm{dim}$ is enhanced by including $V_{b_1}$,~\cite{Tamura} which results in DMI, because $V_{b_1}$ rapidly increases the tendency for localizing one hole on each dimer.
At the same time, AF long-range order appears due to the weak hopping frustration.
This AF DMI is widely observed in $\kappa$-(ET)$_2$X, including the deuterated $\kappa$-(ET)$_2$Cu[N(CN)$_2$]Br.~\cite{Kanoda,Miyagawa}

As $V_{b_1}/U$ increases further, electrons further avoid each other and PCOI is stabilized.
Indeed, the PCOI configuration can avoid $V_{b_1}$,$V_{b_2}$, and $V_p$, at the expense of the energy loss of $V_q$. 
The AF long-range order also appears in PCOI and its spin configuration changes from PCOI-1 to PCOI-2 within the PCOI,~\cite{Naka1}
as schematically shown in Figs.~\ref{fig2}(c) and~\ref{fig2}(d). 
Although these PCOIs have not been observed experimentally, it is proposed that the dielectric anomalies in $\kappa$-(ET)$_2$Cu$_2$(CN)$_3$ originate from the polar charge fluctuation.~\cite{Hotta,Naka1}

When $V_{b_1}/U>0.55$, 3-fold COM appears.
There, the double occupancy of electrons within molecules, namely, the energy loss in the $U$ term, increases to fully avoid the intersite Coulomb interactions.
Since the energy gap of 3-fold CO opens far from the Fermi level, the system remains metallic and has no spin orders. 
Such COM state is often found in 3/4-filled EHM with different geometries.~\cite{Watanabe1,Nishimoto,Naka2}

Finally, SC appears next to the DMI and 3-fold COM phases.
The region of SC is limited for the parameters used here.~\cite{note2}
Nevertheless, finite condensation energy is widely observed in the phase diagram enclosed with red dotted curves in Fig.~\ref{fig2}(a).
The gap function has four nodes and changes the sign four times along the Fermi surface.~\cite{note3}
This structure of the gap function has been discussed in previous studies.~\cite{TWatanabe,Kuroki1,Sekine,Guterding2,Powell}
However, the fundamental difference from these previous studies is that 
the sufficient intersite Coulomb interactions are indispensable for the appearance of the SC.
In fact, the SC is not stabilized along the Mott transition line (phase boundary between PM and DMI) but near CO instabilities.
This strongly indicates the importance of the intersite Coulomb interactions.

\begin{figure}
\begin{center}
\includegraphics[width=7.5cm]{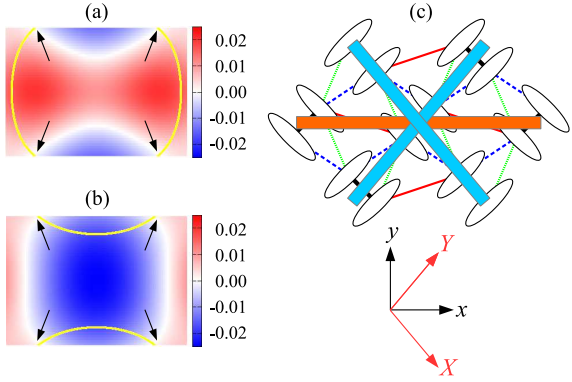}
\caption{\label{fig3}
(Color online) Contour plots for the SC gap function of (a) band 1 ($\Delta^1$) and (b) band 2 
($\Delta^2$) in the first Brillouin zone for $(U/t_{b_1},V_{b_1}/U)=(7.0,0.55)$.
The yellow solid curves represent the Fermi surface for each band.
The arrows indicate node points on the Fermi surface.
(c) Schematic real space pairing of extended-$s$+$d_{XY}$-wave SC. 
The $x$- and $y$-axes are parallel and perpendicular to horizontal bonds (orange solid bar), respectively, whereas
the $X$- and $Y$-axes are parallel to two diagonal bonds (blue solid bars) and in the similar coordinate system used for high-$T_c$ cuprates.
}
\end{center}
\end{figure}

Let us comment on the system size dependence of our result.
The 3-fold COM is properly treated only with $L=12,24,36,\cdots$, where the system size fits the unit cell of 3-fold COM.
We have checked that the phase boundary in Fig.~\ref{fig2} is almost unchanged for $L=12$ and 24 except for SC, while $L=36$ is beyond the computational limit.
The size dependences of the phase boundary and condensation energy of SC are discussed in Supplemental Material.~\cite{sm}

Here, we show the details of the SC gap function.
Figures~\ref{fig3}(a) and~\ref{fig3}(b) show the gap functions of band 1 ($\Delta^1$) and band 2 ($\Delta^2$) for $(U/t_{b_1},V_{b_1}/U)=(7.0,0.55)$, respectively.
$\Delta^1$ ($\Delta^2$) is positive (negative) around the Fermi energy, 
and therefore the gap function has four nodes on the Fermi surface and changes the sign four times.
The explicit form is given as
\begin{align}
\Delta^{\alpha}&=\Delta^{\alpha}_1\left[\cos\left(\frac{1}{2}k_xR_x+k_yR_y\right)
+\cos\left(\frac{1}{2}k_xR_x-k_yR_y\right)\right] \notag \\
&+\Delta^{\alpha}_2\cos k_xR_x \notag \\
&+\cdots \text{(up to 22nd neighbors)}, \label{gap}
\end{align}
where $\alpha\,(=1,2)$ denotes a band index.~\cite{note4}

The two terms in the first line of Eq.~(\ref{gap}) correspond to the pairing in the diagonal directions in real space as shown in Fig.~\ref{fig3}(c) with blue solid bars.
These two pairings have the same sign and can be regarded as extended-$s$ pairing.
It is convenient to use another coordinate system, $X$ and $Y$, diagonal to the $xy$ coordinate, 
as shown in Fig.~\ref{fig3}(c) and Supplemental Material.~\cite{sm}
The $XY$ coordinates are similar to those used in, e.g., the $t$-$J$ model for high-$T_c$ cuprates~\cite{Dagotto}
and in the dimer Hubbard model for $\kappa$-(ET)$_2$X,~\cite{Kino2,Kondo,Schmalian}
where $d_{X^2-Y^2}$ pairing is stabilized. 
Note that this is distinct from the extended-$s$ pairing found here.
Furthermore, the gap functions change the sign between different bands, namely, $\operatorname{sgn}\Delta^1_1=-\operatorname{sgn}\Delta^2_1$.
In this respect, the pairing symmetry can also be referred to as $s_{\pm}$, similar to that of iron-based SC.~\cite{Mazin,Kuroki2}

The $\Delta^{\alpha}_2$ term in the second line of Eq.~(\ref{gap}) corresponds to the pairing in the horizontal direction as shown in Fig.~\ref{fig3}(c) with the orange solid bar.
In the $XY$ coordinates, it is denoted as $d_{XY}$ pairing, which is expected in the dimer Hubbard model with quasi one-dimensional anisotropy.~\cite{TWatanabe,Powell}
In our calculation, the absolute values of $\Delta^{\alpha}_1$ and $\Delta^{\alpha}_2$ are comparable whenever the SC has finite condensation energy.
Therefore, the symmetry of the gap function can be regarded as ``extended-$s$+$d_{XY}$''-wave using the $XY$ coordinates~\cite{Powell,Guterding1,Guterding2}
(extended-$s$+$d_{x^2-y^2}$-wave using the original coordinates).
If the $d_{X^2-Y^2}$ pairing is formed, there would appear a node along the horizontal direction and thus no pairing contributed along this direction.
Note that the symmetry of the gap function does not change within the parameter space investigated here, including the region where SC is metastable
[between the red dotted curves in Fig.~\ref{fig2}(a)].

\begin{figure}
\begin{center}
\includegraphics[width=7.5cm]{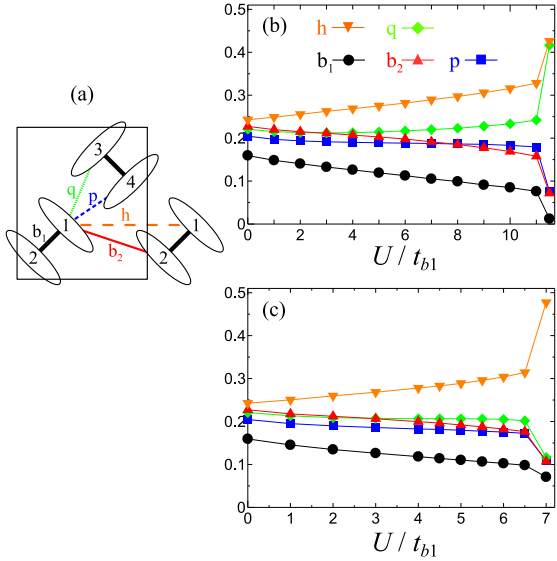}
\caption{\label{fig4}
(Color online) (a) Configuration of bonds up to fifth neighbor.
$U/t_{b_1}$ dependence of charge correlation $\bigl<n_in_j\bigr>$ of each bond indicated in (a) connecting molecular sites $i$ and $j$
for $V_{b_1}/U=0.5$ (b) and $0.6$ (c).
Normal state without any long-range orders is assumed.
}
\end{center}
\end{figure}

To see the effect of the intersite Coulomb interactions more explicitly,
we calculate the charge correlation $\bigl<n_in_j\bigr>$ up to the fifth-neighbor bond 
[see Fig.~\ref{fig4}(a)], for the normal state without any long-range orders ($\{D\}$, $\{M\}$, and $\{\Delta\}$ are all set to zero).
For $V_{b_1}/U<0.55$, the charge correlation of $q$ and $h$ bonds is enhanced and others are suppressed by the increase in $U$, as shown in Fig.~\ref{fig4}(b).
This is consistent with the tendency toward PCOI [see Figs.~\ref{fig2}(c) and~\ref{fig2}(d)].
At $U/t_{b_1}\sim11.5$, $\bigl<n_in_j\bigr>$ shows an abrupt change and the system enters deep inside the PCOI region.
As shown in Fig.~\ref{fig2}(a), the region where finite SC condensation energy is found is located around $7<U/t_{b_1}<11.5$:
This is where such charge correlation is enhanced but before the abrupt change.
Both $q$ and $h$ bonds contribute to the singlet pairing, and then the SC correlation is enhanced there.
However, the $q$ bond also contributes to the AF correlation in DMI and PCOI [see Figs.~\ref{fig2}(b)--\ref{fig2}(d)], thus resulting in the competition among SC, DMI, and PCOI.

On the other hand, for $V_{b_1}/U>0.55$, only the charge correlation of the $h$ bond is enhanced toward 3-fold COM, as shown in  Fig.~\ref{fig4}(c).
Again, in the region where finite SC condensation energy is found ($4<U/t_{b_1}<7$), the charge correlation is enhanced and then contributes to the singlet formation in the horizontal direction
(orange solid bar in Fig.~\ref{fig3}(c)).
However, no AF ordered phases are stabilized in this case.
Thus, the different behaviors in the charge correlation for $V_{b_1}/U$ below and above 0.55 cause a much different phase competition.
Nevertheless, the SC is ubiquitously stabilized when the enhanced charge correlation is observed toward both PCOI and 3-fold COM,
indicating that the intersite Coulomb interaction is an important factor for stabilizing the SC.
A weak-coupling approach with intersite Coulomb interactions supports such results.~\cite{Sekine}

The symmetry of the SC found here seems counterintuitive at a glance because the hopping frustration is rather weak and therefore $d_{X^2-Y^2}$ pairing is naively expected.
Indeed, extended-$s$+$d_{XY}$ and $d_{X^2-Y^2}$ pairings are suggested to compete with each other when the intersite Coulomb interactions are absent.~\cite{Kuroki1,Guterding2}
According to our result, SC is not stabilized without the intersite Coulomb interactions because the charge degree of freedom is still active within the dimer, even with large $U$.
The introduction of intersite Coulomb interactions enhances the instability of SC and finally the SC occurs near the charge-ordered phases.
The intersite Coulomb interactions favor extended-$s$+$d_{XY}$ pairing by enhancing both diagonal and horizontal pairing as discussed above, and it results in the suppression of $d_{X^2-Y^2}$.

Finally, we compare our result with experiments.
According to the first-principles band calculation, members of $\kappa$-(ET)$_2$X exhibiting AF DMI and SC have similar tight-binding parameters.~\cite{Koretsune2}
Therefore, we expect that the phase diagram in Fig.~\ref{fig2}(a) provides a general view of $\kappa$-(ET)$_2$X, except for the highly frustrated member X=Cu$_2$(CN)$_3$ exhibiting spin-liquid behavior.
We argue that these compounds locate in our phase diagram around $U/t_{b_1}$=5-10 (1-2 eV) and $V_{b_1}/U$=0.5-0.55, where several phases are competing with each other.
Indeed, the AF DMI is located on the large-$U/t_{b_1}$ (low-pressure) side and the SC is located on the small-$U/t_{b_1}$ (high-pressure) side,~\cite{note2}
which is consistent with the experimental phase diagram.~\cite{Kanoda}
Although the symmetry of the SC is still controversial,~\cite{Izawa,Ichimura,Guterding1,Elsinger,Taylor,Malone,Milbradt,Oka} most of the experiments support unconventional $d$-wave symmetry.
Some of them~\cite{Izawa,Ichimura} suggest the extended-$s$+$d_{x^2-y^2}$-wave, the same symmetry 
obtained in our study, and the node positions observed experimentally are also comparable to our results.
However, other experiments~\cite{Malone} suggest the $d_{xy}$-wave similar to that of the high-$T_c$ cuprates.
For further discussions, the multi-band effect on the SC should be considered when the experimental results are analyzed because most of the analyses are based on a simpler picture obtained from single-band models.

To summarize, we have studied the 3/4-filled EHM to examine the phase competition and the mechanism of SC in $\kappa$-(ET)$_2$X.
We have shown that the cooperation between the on-site and intersite Coulomb interactions induces various phases such as DMI, PCOI, 3-fold COM, and SC.
The symmetry of SC is found to be the ``extended-$s$+$d_{x^2-y^2}$''-wave and the charge correlation enhanced toward the charge-ordered phases is a key factor for stabilizing the SC.
Our results demonstrate the importance of the intradimer charge degree of freedom for the unified description of various competing phases in $\kappa$-(ET)$_2$X
as well as molecular materials with 3/4-filled band in general.
Note also that the spin-liquid DMI is another long-standing issue in $\kappa$-(ET)$_2$X.
Whether the 3/4-filled EHM can describe the spin-liquid DMI is a highly interesting issue and is left for a future problem.

\section*{Acknowledgments}

The authors thank T. Koretsune and T. Shirakawa for useful discussions.
The computation has been carried out using the facilities of the Supercomputer Center, Institute for Solid State Physics, University of Tokyo. 
This work has been supported by JSPS KAKENHI Grant Nos. 26800198, 26287070, 26400377, and 16H02393 and in part by RIKEN iTHES Project and Molecular Systems.


\begin{thebibliography}{99}
\bibitem{note1} We have checked that the obtained phase diagram is qualitatively the same when $V_{b_2}=V_p=V_q$ is assumed.
Our result will be unchanged as long as $V_{b_2}$, $V_p$, and $V_q$ have values close to each other.
\bibitem{note2} Generally, the VMC method overestimates the AF insulating region over normal metal and SC, although it is much improved compared with a mean-field-type approximation.
\bibitem{note3} Depending on the value of Coulomb interactions, the node position is slightly moved from the edge of the first Brillouin zone and can intersect the Fermi surface of band 2 (See Fig.~\ref{fig3}).
However, the number of nodes on the Fermi surface and the symmetry of the gap function do not change.
\bibitem{note4} We optimize the real space pairing up to 22nd neighbors and find that the overall feature of the gap function is determined within the fourth neighbor, i.e., $\Delta^{\alpha}_m$ for $m\leq 4$.
\end{thebibliography}
\end{document}


\title{
Supplemental Material for\\
Phase Competition and Superconductivity in $\kappa$-(BEDT-TTF)$_2$X:\\
Importance of Intermolecular Coulomb Interactions
}


\author{Hiroshi Watanabe$^{1}$}
\author{Hitoshi Seo$^{2,3}$}
\author{Seiji Yunoki$^{2,3,4}$}
\affiliation{
$^1$Waseda Institute for Advanced Study, Waseda University, Shinjuku, Tokyo 169-8050, Japan\\
$^2$RIKEN Center for Emergent Matter Science (CEMS), Wako, Saitama 351-0198, Japan\\
$^3$RIKEN, Wako, Saitama 351-0198, Japan\\
$^4$RIKEN Advanced Institute for Computational Science (AICS), Kobe, Hyogo 650-0047, Japan
}



\maketitle

Here we describe the trial wave function $|\Psi\rangle$ employed in the variational Monte Carlo calculations for the ground state of the 3/4-filled EHM studied in the main text.
The Fermi surface and energy dispersion of the noninteracting tight-binding energy band for $\kappa$-(ET)$_2$Cu[N(CN)$_2$]Br are shown in Fig.~S1.
The one-body part $\left|\Phi\right>$ for PM, DMI, and PCOI-1 can be obtained by diagonalizing the one-body Hamiltonian,

\begin{figure}[t]
\begin{center}
\includegraphics[width=0.7\hsize]{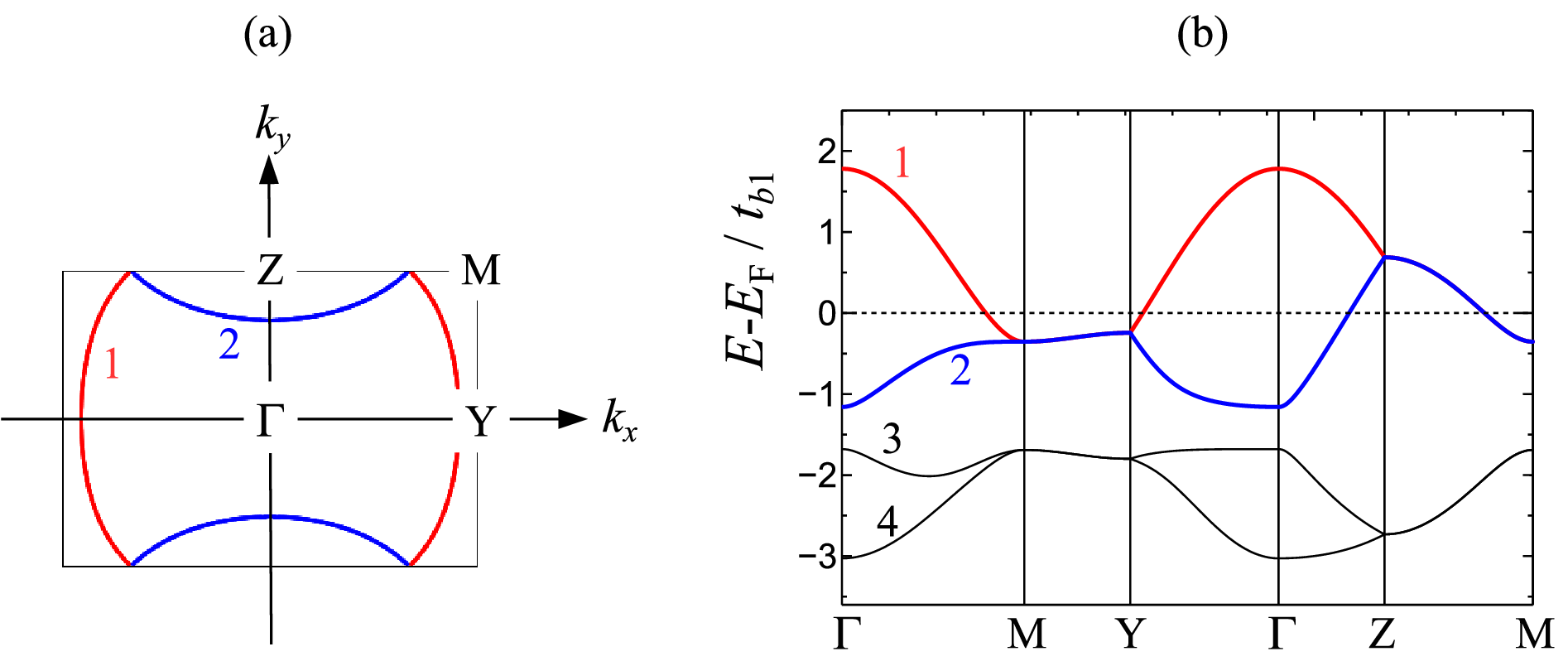}
\caption{
(Color online) (a) Fermi surface and (b) energy dispersion of the noninteracting tight-binding energy band for $\kappa$-(ET)$_2$Cu[N(CN)$_2$]Br. The tight-binding parameters are given in the main text.
High symmetry points are $\Gamma$(0,0), Y($\pi/R_x$), Z($\pi/2R_y$), and M($\pi/R_x,\pi/2R_y$).
Numbers in (a) and (b) denote the band index $\alpha\,(=1,2,3,4)$, and only bands 1 and 2 cross the Fermi energy $E_{\text{F}}$.
}
\end{center}
\end{figure}

\begin{align}
\tilde{H}_0&=\sum_{\bm{k}\sigma}
        \left( c_{\bm{k}1\sigma}^{\dg}, c_{\bm{k}2\sigma}^{\dg},  
   c_{\bm{k}3\sigma}^{\dg}, c_{\bm{k}4\sigma}^{\dg}\right)  
 \begin{pmatrix}
   -\sigma M^z_1 & T^*_{21} & T^*_{31} & T^*_{41} \\ 
   T_{21} & -\sigma M^z_2 & T^*_{32} & T^*_{42} \\
   T_{31} & T_{32} & \sigma M^z_3 & T^*_{43} \\
   T_{41} & T_{42} & T_{43} & \sigma M^z_4
 \end{pmatrix}
 \begin{pmatrix}
  c_{\bm{k}1\sigma} \\ c_{\bm{k}2\sigma} \\ 
  c_{\bm{k}3\sigma} \\ c_{\bm{k}4\sigma}    
 \end{pmatrix} \label{eq1} \tag{S1} \\
      &=\sum_{\bm{k}\alpha\sigma}\tilde{E}_{\alpha}(\bm{k})a^{\dg}_{\bm{k}\alpha\sigma}a_{\bm{k}\alpha\sigma}, \tag{S2} \label{eq:qp}
\end{align}
with the hopping matrix elements given as
\begin{align}
T_{21}&=-\tilde{t}_{b_1}\text{e}^{-i(k_x\delta_x+k_y\delta_y)}-\tilde{t}_{b_2}\text{e}^{i\{k_x(R_x-\delta_x)-k_y\delta_y)\}}, \tag{S3} \\
T_{31}&=-\tilde{t}_q\left[\text{e}^{i\{k_x(R_x/2-\delta_x)+k_yR_y)\}}+\text{e}^{i\{k_x(R_x/2-\delta_x)-k_yR_y)\}}\right] \notag \\
      &=-2\tilde{t}_q\text{e}^{ik_x(R_x/2-\delta_x)}\cos k_yR_y, \tag{S4} \\
T_{32}&=-2\tilde{t}_p\text{e}^{-ik_y(R_y-\delta_y)}\cos \frac{1}{2}k_xR_x, \tag{S5} \\
T_{41}&=-2\tilde{t}_p\text{e}^{ik_y(R_y-\delta_y)}\cos \frac{1}{2}k_xR_x, \tag{S6} \\
T_{42}&=-2\tilde{t}_q\text{e}^{-ik_x(R_x/2-\delta_x)}\cos k_yR_y, \tag{S7} \\
T_{43}&=-\tilde{t}_{b_1}\text{e}^{i(k_x\delta_x-k_y\delta_y)}-\tilde{t}_{b_2}\text{e}^{-i\{k_x(R_x-\delta_x)+k_y\delta_y)\}}, \tag{S8}
\end{align}
where $c^{\dg}_{\bm{k}m\sigma}$ ($c_{\bm{k}m\sigma}$) is a creation (annihilation) operator of electron at molecule $m$(=1--4),
as indicated in Fig.~1 (a) in the main text, with momentum $\bm{k}$ and spin $\sigma\,(=\uparrow,\downarrow)$.
$M^z_m$ is a variational parameter which induces the staggered AF long-range order aligned to $z$-axis for molecule $m$;
$M^z_1=M^z_2=M^z_3=M^z_4=0$ for PM, $M^z_1=M^z_2=M^z_3=M^z_4\neq 0$ for DMI ($z$-DMI), and $M^z_1=M^z_3\neq M^z_2=M^z_4$ for PCOI-1 ($z$-PCOI-1).
$a^{\dg}_{\bm{k}\alpha\sigma}$ ($a_{\bm{k}\alpha\sigma}$) in Eq.~(\ref{eq:qp})
is a creation (annihilation) operator of quasiparticle in band $\alpha$ with momentum $\bm{k}$ and spin $\sigma\,(=\uparrow,\downarrow)$,
obtained by diagonalizing Eq.~(\ref{eq1}), and $\tilde{E}_{\alpha}(\bm{k})$ is the corresponding quasipaticle energy.
The unit cell for PCOI-2 contains 8 molecules, twice as large as the original one, as shown in Fig.~2 (d) in the main text.
Therefore, $\left|\Phi\right>$ for PCOI-2 is obtained by diagonalizing 8$\times$8 matrix similar to Eq.~(\ref{eq1}).

We also consider the phase with the staggered AF moment aligned to $x$-axis: $x$-DMI, $x$-PCOI-1, and $x$-PCOI-2.
The corresponding Weiss field can be expressed by a term such as $M^x_m\sum_{\bm{k}}(c^{\dg}_{\bm{k}m\ua}c_{\bm{k}m\da}+c^{\dg}_{\bm{k}m\da}c_{\bm{k}m\ua})$,
where $M_m^x$ is a variational parameter.
The dimension of the matrix is doubled because of the mixing of up- and down-spin components: 8$\times$8 for $x$-DMI and $x$-PCOI-1, and 16$\times$16 for $x$-PCOI-2.

We have found that the variational energy of $z$-DMI is always lower than that of $x$-DMI.
On the other hand, the variational energy of $x$-PCOI-1 ($x$-PCOI-2) is always lower than that of $z$-PCOI-1 ($z$-PCOI-2).
The difference between $z$- and $x$-ordered phases is due to the presence of the spin Jastrow factor $P_{\text{J}_{\text{s}}}=\exp[-\sum_{i,j}v^{\text{s}}_{ij}s^z_is^z_j]$, which breaks the spin rotational symmetry of the trial wave function. 
Note however that we have not described the direction of the staggered AF moment in the main text.

For 3-fold COM, the unit cell is three times as large as the original one and contains 12 molecules, as shown in Fig.~2 (e) in the main text.
Therefore, $\left|\Phi\right>$ for 3-fold CO is obtained by diagonalizing 12$\times$12 matrix.
The corresponding Weiss field is $D_m\sum_{\bm{k}\sigma}c^{\dg}_{\bm{k}m\sigma}c_{\bm{k}m\sigma}$, which induces charge disproportionation within the unit cell through the variational parameters $D_m$ for $m=1,2,\cdots,12$.

Finally, $\left|\Phi\right>$ for SC is obtained by diagonalizing the BCS-type mean-field Hamiltonian,
\begin{widetext}
\begin{equation}
\tilde{H}_{\mathrm{BCS}}=\sum_{\bm{k}}
    \left(a_{\bm{k}1\ua}^{\dg}, a_{\bm{k}2\ua}^{\dg}, a_{\bm{k}3\ua}^{\dg}, a_{\bm{k}4\ua}^{\dg}
   a_{-\bm{k}1\da}, a_{-\bm{k}2\da}, a_{-\bm{k}3\da}, a_{-\bm{k}4\da}\right)
  \begin{pmatrix}
   \xi_1 & 0 & 0 & 0 & \Delta^1 & 0 & 0 & 0\\
   0 & \xi_2 & 0 & 0 & 0 & \Delta^2 & 0 & 0\\
   0 & 0 & \xi_3 & 0 & 0 & 0 & \Delta^3 & 0\\
   0 & 0 & 0 & \xi_4 & 0 & 0 & 0 & \Delta^4\\
   \Delta^1 & 0 & 0 & 0 & -\xi_1 & 0 & 0 & 0\\
   0 & \Delta^2 & 0 & 0 & 0 & -\xi_2 & 0 & 0\\
   0 & 0 & \Delta^3 & 0 & 0 & 0 & -\xi_3 & 0\\
   0 & 0 & 0& \Delta^4 & 0 & 0 & 0 & -\xi_4
 \end{pmatrix}
 \begin{pmatrix}
a_{\bm{k}1\ua} \\ a_{\bm{k}2\ua} \\ a_{\bm{k}3\ua} \\ a_{\bm{k}4\ua} \\  
   a_{-\bm{k}1\da}^{\dg}\\ a_{-\bm{k}2\da}^{\dg}\\ a_{-\bm{k}3\da}^{\dg} \\ a_{-\bm{k}4\da}^{\dg}\end{pmatrix}, \tag{S9}
\end{equation}
\end{widetext}
where $\Delta^{\alpha}$ denotes gap function for band $\alpha$ and $\xi_{\alpha}=\tilde{E}_{\alpha}(\bm{k})-\tilde{\mu}$ is a quasiparticle energy measured from the renormalized chemical potential $\tilde{\mu}$.
$\Delta^{\alpha}$ is constructed from the real space pairing up to 22nd neighbors.
Since the band 3 and 4 are located much below the Fermi energy, their contribution to the pairing is negligible and therefore we set $\Delta^3=\Delta^4=0$.
We have found that the condensation energy becomes finite only for the extended-$s$+$d_{x^2-y^2}$-wave among several candidates of the gap function.
The explicit form of $\Delta^1$ and $\Delta^2$ is given in Eq.~(2) in the main text.

In the main text, we have used two different coordinate systems, $(x, y)$ and $(X, Y)$, which are compared in Fig.~S2.
Using $XY$ coordinates, the SC gap function $\Delta^{'\alpha}$ which corresponds to Eq.~(2) in the main text is expressed as
\begin{equation}
\Delta^{'\alpha}=\Delta^{'\alpha}_1\left[\cos(k_XR_X)+\cos(k_YR_Y)\right]+\Delta^{'\alpha}_2\cos(k_XR_X+k_YR_Y)+\cdots \text{(up to 22nd neighbors)}. \tag{S10}
\end{equation}
In this coordinate system, the symmetry is extended-$s$+$d_{XY}$-wave type, as discussed in the main text.
For $d_{X^2-Y^2}$-wave type, the dominant part of the SC gap function is $\sim\cos(k_XR_X)-\cos(k_YR_Y)$, which is the same form of the SC gap function generally discussed in high-$T_c$ cuprates.

\begin{figure}[t]
\begin{center}
\includegraphics[width=0.9\hsize]{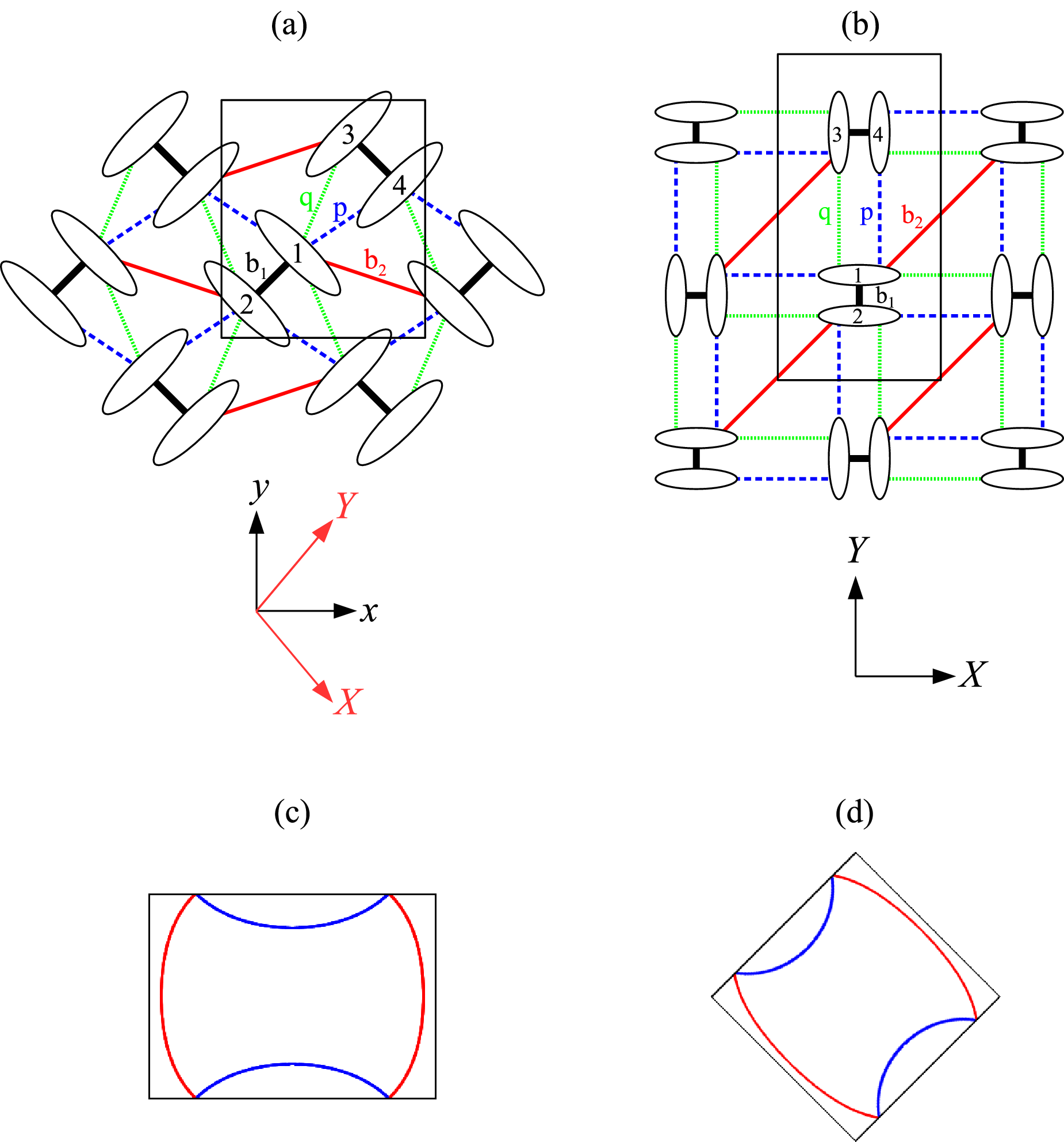}
\caption{
(Color online) Comparison between two different coordinate systems.
Lattice structures in (a) ($x,y$) and (b) ($X,Y$) planes.
Black rectangles indicate unit cells with (a) $R_x\times 2R_y$ and (b) $R_X\times 2R_Y$.
Fermi surfaces are also shown in (c) ($k_x,k_y$) and (d) ($k_X,k_Y$) planes. 
}
\end{center}
\end{figure}

\begin{figure}[t]
\begin{center}
\includegraphics[width=0.5\hsize]{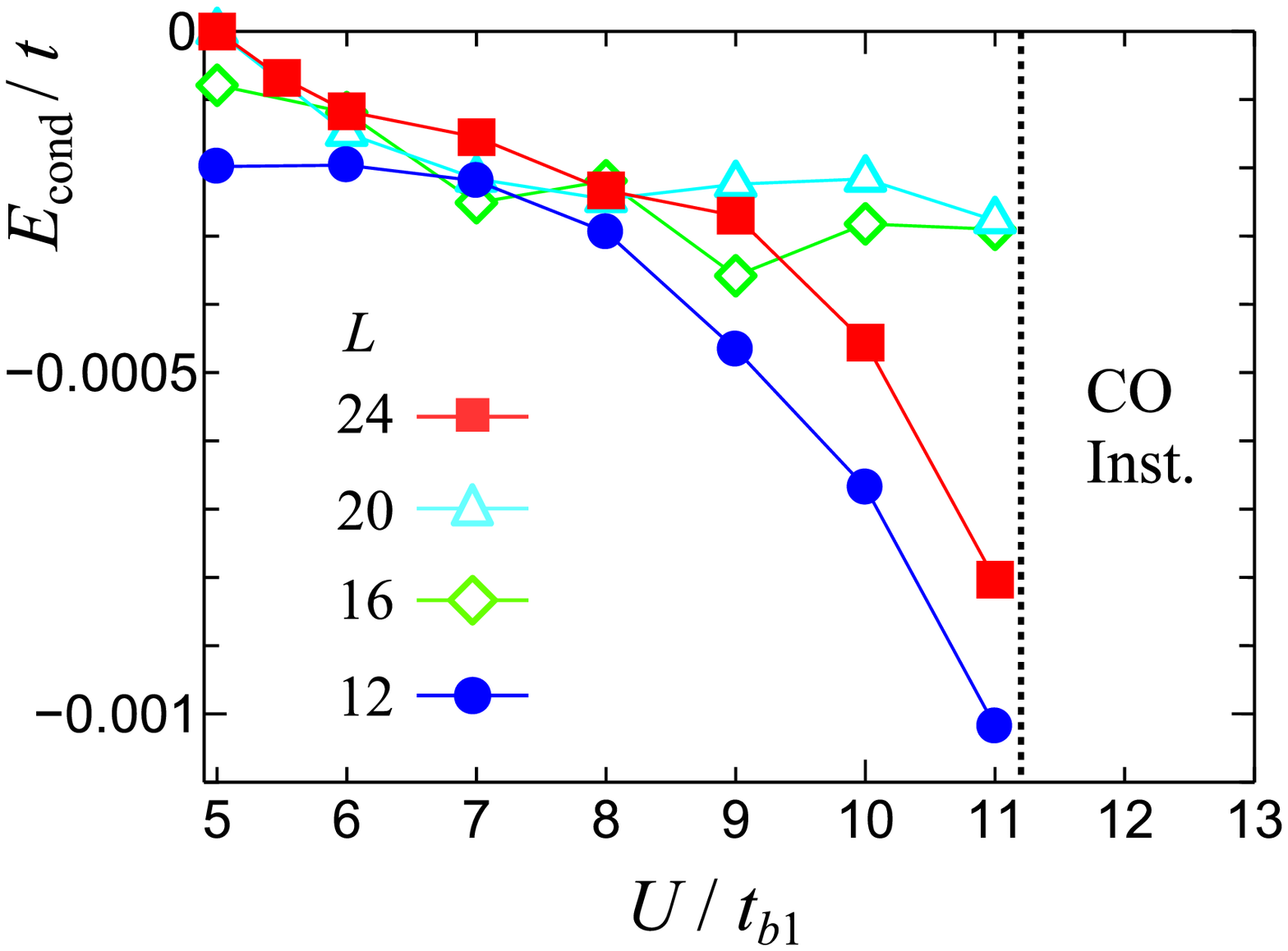}
\caption{\label{figS3}
(Color online) Size dependence of the condensation energy of SC (per molecular site) for $V_{b_1}/U=0.52$. CO Inst. denotes the charge-order instability of 3-fold type.}
\end{center}
\end{figure}

Finally, we disucuss the system size dependence of the result for SC in the main text.
Figure~\ref{figS3} shows the $U$ dependence of the condensation energy (per molecular site) $E_{\text{cond}}$, for $V_{b_1}/U=0.52$, for several different $L$.
$E_{\text{cond}}$ is estimated as the energy gain compared with normal state without long-range order.
For $L=12$ and 24, $E_{\text{cond}}$ is clearly enhanced toward the charge-order instability of 3-fold type.
On the other hand, such enhancement is not observed for $L=16$ and 20, where the system size does not fit the unit cell of 3-fold charge-ordered phase and the charge instability is not appropriately included.
These results indicate that the 3-fold type charge fluctuation is important for the SC.
Due to the finite size effect, the region where $E_{\text{cond}}$ is finite for $L=24$ is smaller than that for $L=12$.
It is difficult to discuss the systematic size dependence (next availabe size is $L=36$ but too much time-consuming) and therefore we use the result for $L=24$ in the main text,
which we believe to be satisfactory for the stability of SC phase.

